\documentclass[floatfix,showpacs,amsmath,amssymb,letterpaper,groupaddresses,superscriptaddress]{article}
\usepackage{graphicx}
\usepackage{latexsym}
\usepackage{amsmath,amssymb}

\usepackage{epsfig}
\usepackage{graphicx}

\usepackage{graphicx}
\usepackage{amssymb,amsmath,times}

\usepackage{color}

\def\be{\begin{equation}}
\def\ee{\end{equation}}
\def\ba{\begin{eqnarray}}
\def\ea{\end{eqnarray}}
\def\la{\langle}
\def\ra{\rangle}
\def\a{\alpha}
\def\nn{\nonumber}

\def\h{\hskip 1cm}

\def\lo{\longrightarrow}

\begin{document}
\begin{center}
{\Large Secure Quantum Carriers for Quantum State Sharing}\\
\vspace{1cm} Vahid Karimipour\footnote{email: vahid@sharif.edu,
corresponding author}\\

Department of Physics, Sharif University of Technology,\\
P.O. Box 11155-9161,\\ Tehran, Iran

\vspace{1cm}

 Milad Marvian \footnote{email:milad.marvian@ee.sharif.edu}
\\

Department of Electrical Engineering, Sharif University of Technology, \\
 Tehran, Iran.

\end{center}

\vspace{5mm}

\begin{abstract}
We develop the concept of quantum carrier and show that messages
can be uploaded and downloaded from this carrier and while in
transit, these messages are hidden from external agents. We
explain in detail the working of the quantum carrier for different
communication tasks, including quantum key distribution,
classical secret and quantum state sharing among a set of $n$
players according to general threshold schemes. The security of
the protocol is discussed and it is shown that only the
legitimate subsets can retrieve the secret messages, the
collaboration of all the parties is needed for the continuous
running of the protocol and maintaining the carrier.

\end{abstract}
\vskip 1cm Keywords: Quantum carrier; Entanglement; Secret sharing; Threshold schemes.\\

\section{Introduction}\label{intro}

Entanglement has been used in many different protocols of quantum
information theory, from teleportation and key distribution to
secret sharing
\cite{hbb,karlsson,tripartite,tyc,contisanders,Exp1,Exp2,bk,zhang2,liy}.
In all these protocols, entanglement is a resource which is
completely consumed by measurements of the parties involved and
should be generated anew for next rounds of protocol. It is true
that generating and maintaining entanglement between several
particles is very difficult. Yet with the developments in
realizing quantum repeaters
\cite{repeaters1,repeaters2,repeaters3}, creating and maintaining
long distance entanglement between stationary quantum systems
becomes more feasible in the (possibly distant) future. It is
thus rewarding to imagine if this entanglement can used in a
different way, that is as a carrier of information, which
modulates and transmits quantum information, in the same way as
carrier waves in classical communication systems carry modulated
messages. In this new application, we can imagine that arbitrary
quantum states are uploaded (entangled) to the carrier by the
sender and downloaded (disentangled) from the carrier by the
receiver(s), in such a way that at the end of the protocol, the
carrier remains intact and ready for use in next rounds. The role
of the quantum carrier and its entanglement with the messages
will be to hide messages from adversaries, hence the term secure
quantum
carrier.\\

This new way for using entanglement was first reported in
\cite{zhang} for quantum key distribution and then was developed
for a simple secret sharing scheme in \cite{bk}. In this paper we
want to develop it further and provide quantum carriers for
secret sharing schemes for general threshold schemes
\cite{cleve,sarvepalli}.  Needless to say, our aim is not to
develop new quantum secret sharing schemes, but to develop a
quantum carrier for distributing in a secure way the already
known quantum secrets among the parties. Our emphasis is thus on
the very concept of secure quantum carrier and
the way it can be used in quantum communication protocols. In the particular context of quantum secret sharing, as we will see it allows us to
generate broader threshold schemes than those of \cite{bar1,bar2}. \\

In particular we have to stress the difference with the works in
\cite{bar1,bar2} where it was shown that graph state formalism
\cite{graph} can act as a framework for unifying some of the
secret sharing protocols, albeit not for general threshold
structures. The idea of \cite{bar1,bar2} was to encode the
secret in some local actions of the dealer on a vertex of a
suitably chosen weighted graph state. Local measurements of the
players on different vertices of this graph, could then reveal
the secret to authorized subsets of players. In this way, it was
shown that threshold schemes of the type $(n,n), (2,3)$ and
$(3,5)$ can be implemented in a unified way for various forms of
channels interconnecting the dealer and the players. Therefore
the works of \cite{bar1,bar2} belong to the same class as in
\cite{hbb}, in which entanglement is fully consumed due to
measurements of the players. \\

It is to be noted that while the idea of a fixed quantum carrier
has an appeal for communication, a price should be paid for its
implementation: it requires a larger number of particles to be
entangled at the beginning and end of the protocol, but at the
end of each round a fixed amount of entanglement remains in the
form of a carrier. Nevertheless, it is worth to develop such a
concept from theoretical side and hope that it will someday
become close to reality.\\

We remark that although we present our analysis for secure
communication of basis states or classical information, the idea
also works for sending arbitrary quantum states. In the simplest protocol discussed in the beginning of the paper we explicitly show this, although we will not
repeat it for other general schemes. \\

The structure of this paper is as follows. In section
\ref{general} we put forward the basic requirements that a
quantum carrier should satisfy, in section \ref{keycarrier} we
explain the basic method in the simplest possible setting, that
is quantum key distribution between two parties. Then in section
\ref{bkcarrier} we briefly explain the use of quantum carrier for
the simplest secret sharing scheme, where one dealer, wants to
share a secret between two different players who have equal right
for retrieving the message collaboratively. For this reason, this
is called a $(2,2)$ secret shring scheme. This is then
generalized to the $(n,n)$ scheme where a secret is to be shared
between $n$ players and all the $n$ players can retrieve the
message collaboratively. Finally in section \ref{kncarrier} we
define the carrier for the $(k,n)$ threshold scheme where any of
the $k$ players can retrieve the secret, although collaboration
of all of the players is needed for the continuous running and
security of the protocol. We end the paper with conclusion and
outlook.

\section{General Considerations On the Quantum
Carrier}\label{general}

Suppose that a quantum carrier has been set up for a specific
communication task, i.e. for a quantum key distribution between
Alice and Bob or a secret sharing scheme, between Alice as the
dealer and Bob and Charlie as the players. This quantum carrier
should have the following properties:

\begin{itemize}
\item There should be simple and local uploading and
downloading operators, so that the legitimate parties can upload
and download messages to or from this carrier.

\item While in transit, the messages should be hidden from
third parties so that no intercept-resend strategy can reveal the
identity of the message.

\item Eve should not be able to entangle herself to the
quantum carrier without being detected by the legitimate
parties. This property is to prevent Eve from conducting more complex attacks.

\end{itemize}

Once such criteria are met, we say that a secure quantum carrier
has been set up for this communication task. In the rest of this
paper we present quantum carriers for various communication tasks.
We should stress again that these requirements are purely from
the theoretical point of view, the main difficulty will obviously
be to maintain the carrier for a long enough time so that it can
be used for passing many quantum states before the entanglement
decays and becomes useless.

\section{Quantum Carrier For Key Distribution}\label{keycarrier}
The first task that we discuss is the simple communication
between two parties, where Alice wants to send a sequence of bits
$0$ and $1$, a classical message, to Bob \cite{zhang}. Alice
encodes the classical bits $0$ and $1$ into states $|0\ra$ and
$|1\ra$ (the eigenbases of the $Z$ operator). The quantum carrier
is
\begin{equation}\label{(1,1)carrier}
|\phi\ra_{a,b}:=\frac{1}{\sqrt{2}}(|0,0\ra+|1,1\ra)_{a,b},
\end{equation}
where $a$ and $b$ refer to the Hilbert spaces of Alice and Bob
respectively. The Hilbert space of the message is denoted by a
number $1$ (since one qubit is being transmitted). The uploading
operator, used by Alice, is a CNOT operator which we denote by
$C_{a,1}$,
\begin{equation}\label{Cam}
  C_{a,1}|i,j\ra_{a,1}=|i,i+j\ra_{a,1}.
\end{equation}
The downloading operator is $C_{b,1}$, i.e. with control port by
Bob and target port the message.\\

Consider now a classical bit $s$ which is encoded to the quantum
state $|s\ra$ and is to be transferred from Alice to Bob. Alice
performs the local operation $C_{a,1}$ on the state
$|\phi\ra_{a,b}|s\ra_1$, turning this state into
\begin{equation}\label{phisab}
  |\phi_s\ra_{a,b,1}=\frac{1}{\sqrt{2}}(|0,0,s\ra+|1,1,s'\ra)_{a,b,1},
\end{equation}
where $s':=s+1\ {\rm mod} \ 2 $.  While in transit the message is
in the state
\begin{equation}\label{rhom}
\rho^s_1=\frac{1}{2}(|s\ra\la s|+|s'\ra\la s'|)=\frac{1}{2}I,
\end{equation}
and hence inaccessible to Eve. At the destination, Bob can
download the message from the carrier by his local operation
$C_{b,1}$, which disentangles the message and leaves the carrier
in its original form, ready for use in the next round. The fact
that Bob downloads exactly the same state which has been uploaded
by Alice is due to the perfect correlation of the states of Alice
and Bob in the carrier. Alice can also use this carrier for
sending quantum states to Bob. Linearity of the uploading and
downloading
operations allows Alice and Bob to entangle and disentangle a quantum state $|\phi\ra=a|0\ra+b|1\ra$ to and from the carrier. \\

To conduct a somewhat complex attacks on the communication, Eve
can entangle herself to the carrier and try to intercept-recend
the message. To do this the only possibility for her entanglement
is
\begin{equation}\label{abe}
  |\phi'\ra_{a,b,e}=|0,0\ra|\xi_0\ra+|1,1\ra|\xi_1\ra,
\end{equation}
where $|\xi_0\ra$ and $|\xi_1\ra$ are two un-normalized states of
Eve's ancilla.  Any other form of entanglement, i.e. one in which
a term like $|0,1\ra|\eta\ra$ is also present in the above
expansion, will destroy the perfect correlation between the
sequence of bits transmitted between Alice and Bob. In case that
the two parties are using the carrier for sending classical bits,
Alice and Bob can publicly compare a subsequence of bits to detect
the presence of Eve's entanglement. In case that they are using
the carrier for sending quantum states, Alice can insert a random
subsequence of basis states into the main stream of states and
ask Bob to publicly announce his results of measurements of these
specific states. This strategy also works in other more complicated schemes presented later, namely the $(n,n)$ and the $(k,n)$ schemes. \\

In order to prevent this type of entanglement, we now use a
property of the carrier (\ref{(1,1)carrier}) which turns out to
be important in all the other forms of quantum carriers that we
will introduce later on. This is the invariance property of the
carrier (\ref{(1,1)carrier}) under Hadamard operations, that is
\begin{equation}\label{hhphi}
  (H\otimes H)|\phi\ra=|\phi\ra.
\end{equation}
At the end of each round, when the message is downloaded and the
carrier is clean, both Alice and Bob act on their share of the
carrier by Hadamard operations. In the absence of Eve, the carrier
will remain the same, however in presence of Eve, (who supposedly
acts on her ancilla by a unitary $U$) the {\it{contaminated}}
(entangled with the ancialla of Eve) carrier (\ref{(1,1)carrier})
will turn out to be
\begin{eqnarray}\label{abe}
  (H\otimes H\otimes U)|\phi'\ra_{a,b,e}&=&|+,+\ra |\eta_0\ra+|-,-\ra|\eta_1\ra\cr
&=&\frac{1}{2}(|0,0\ra +|1,1\ra)(|\eta_0\ra+|\eta_1\ra) \cr &+&
\frac{1}{2}(|0,1\ra +|1,0\ra)(|\eta_0\ra-|\eta_1\ra),
\end{eqnarray}
where $|\eta_0\ra=U|\xi_0\ra$ and $|\eta_1\ra=U|\xi_1\ra$. The
second term in the carrier will certainly introduce
anti-correlations into the basis states communicated between Alice
and Bob, unless $|\eta_0\ra=|\eta_1\ra$ and hence
$|\xi_0\ra=|\xi_1\ra$ which means that Eve cannot entangle
herself to the carrier.

\section{Quantum Carrier For $(2,2)$ Secret
Sharing}\label{bkcarrier}
In this scheme, Alice wants to share a secret with Bob and
Charlie so that they can retrieve the message only by their
collaboration. The first quantum protocol for this scheme was
designed in \cite{hbb} where it was shown that measurements of a
GHZ state in random bases by the three parties can enable them to
share a random secret key. The secure carrier for this protocol
was first developed in \cite{bk}.  Its characteristic feature is
that two types of carriers, should be used which are turned into
each other by the Hadamard operations. The two carriers are
\begin{equation}\label{G}
  |\phi_{\rm odd}\ra = \frac{1}{\sqrt{2}}(|000\ra+|111\ra),
\end{equation}
used in the odd rounds 1, 3, 5, $\cdots$ and
\begin{equation}\label{E}
 |\phi_{\rm even}\ra:=  \frac{1}{2}(|000\ra+|011\ra+|101\ra+|110\ra),
\end{equation}
used in the even rounds, 2, 4, 6, $\cdots.$ The two types of
carriers are turned into each other by the local Hadamard action
of the players at the end of each round,
\begin{equation}\label{HE}
  H^{\otimes 3}|\phi_{\rm odd}\ra=|\phi_{\rm even}\ra, \h  H^{\otimes 3}|\phi_{\rm even}\ra=|\phi_{\rm odd}\ra.
\end{equation}
This property is crucial in checking the security of the protocol
and detection of a Eve attempts who may entangle herself with the
carrier and intercept the secret bits.\\
In the odd and even rounds, the secret bit $s$ is encoded
differently as
\begin{eqnarray}\label{sodd}
&&|\overline{s}_{\rm odd}\ra:=|s,s\ra\cr &&|\bar{s}_{\rm
even}\ra=\frac{1}{\sqrt{2}}(|s,0\ra+|s',1\ra)=\frac{1}{\sqrt{2}}(|+,+\ra+(-1)^s|-,-\ra),
\end{eqnarray}
where $s'=s+1\ {\rm mod}\  2 $. While in the odd rounds, the
receivers each receive a copy of the sent bit, in the even rounds
they need each other collaboration for its retrieval. Therefore Alice can use the odd rounds to put random stray bits and put the message bits in the even rounds. \\
In our opinion this property of the protocol, that is, a rate of
one-half in sending message bits is analogous to discarding
one-half of the measured bits in measurement-based protocols.
However the bonus here is that Alice can send pre-determined
non-random messages.\\

Note that in both even and odd rounds the carrier can be written
as
\begin{equation}\label{carrieruniform}
  |\phi\ra_{a,b,c}:=\frac{1}{\sqrt{2}}\sum_{q=0}|q\ra_a|\overline{q}\ra_{b,c},
\end{equation}
where we have dropped the subscripts even and odd to stress the
uniformity. The running of the protocol, i.e. the uploading and
downloading operations, are based on the following readily
verifiable identities which we state for odd and even rounds separately without writing the subscripts "odd" and "even" explicitly:\\

For odd rounds:
\begin{eqnarray}\label{identitiesodd}
C_{a,1}C_{a,2}|q\ra_a|\overline{s}\ra_{1,2}&=&|q\ra_a|\overline{q+s}\ra_{1,2}\cr
C_{b,1}C_{c,2}|\overline{q}\ra_{b,c}|\overline{s}\ra_{1,2}&=&
|\overline{q}\ra_{b,c}|\overline{q+s}\ra_{1,2}.
\end{eqnarray}
For even rounds:
\begin{eqnarray}\label{identitieseven}
C_{a,1}|q\ra_a|\overline{s}\ra_{1,2}&=&|q\ra_a|\overline{q+s}\ra_{1,2}\cr
C_{b,1}C_{c,2}|\overline{q}\ra_{b,c}|\overline{s}\ra_{1,2}&=&
|\overline{q}\ra_{b,c}|\overline{q+s}\ra_{1,2}.
\end{eqnarray}

Equations (\ref{carrieruniform}, \ref{identitiesodd} and
\ref{identitieseven}) show how the encoded secret can be
downloaded by Alice and downloaded by Bob and Charlie in
different rounds. In the odd rounds, the uploading operator is
simply $C_{a,1}C_{a,2}$, and in the even rounds it is $C_{a,1}$.
In both types of rounds the downloading operator is
$C_{b,1}C_{c,2}$. These string of operators, that is, uploading,
carrying and downloading is depicted as follows:

\begin{equation}\label{diagram}
  |\phi\ra_{a,b,c}|\overline{s}\ra_{1,2}\lo {\rm upload} \lo
  \frac{1}{\sqrt{2}}\sum_{q=0}^1|q,\overline{q},\overline{q+s}\ra_{a,b,c,1,2}\lo
  {\rm download} \lo  |\phi\ra_{a,b,c}|\overline{s}_{1,2}\ra
\end{equation}

The above equation shows that any state
$|\chi\ra=\sum_{s}\a_s|s\ra$ can be encoded as
$|\overline{\chi}\ra=\sum_{s}\a_s|\overline{s}\ra$ and
transferred by the same operations. So this protocol can also be
used for quantum state sharing in a secure way.  The problems of
security of the carriers and the impossibility of Eve's
entanglement with them, has been analyzed in detail in \cite{bk}.
The main points are that i) the secret state is transferred from
Alice to Bob and Charlie in a mixed state and hence carries no
information to outsiders, and ii) the carriers in even and odd
rounds are turned into each other by local Hadamard actions of
Alice, Bob and Charlie, a property which is possible only in the
absence of any entanglement with Eve. Any entanglement will have
a detectable trace on a substring of transferred states, which
will be used to detect the presence of Eve \cite{bk}.

\section{Quantum carrier for $(n,n)$ secret and state
sharing}\label{nncarrier}

The previous protocol can be generalized to the $(n,n)$ scheme,
where all the $n$ players should retrieve the secret
collaboratively. In this case the encoding of a bit $s$ to
quantum states for odd and even rounds is
\begin{eqnarray}\label{encodingnn}
|\overline{s}_{\rm odd}\ra&=&|s\ra^{\otimes n}=|s,s,\cdots s\ra\cr
|\overline{s}_{\rm even}\ra&=& \frac{1}{\sqrt{2}}(|+,+,\cdots
+\ra+(-1)^s|-,-,\cdots -\ra)
\end{eqnarray}
Therefore $|\overline{0}_{\rm even}\ra$ is an even parity state,
and $|\overline{1}_{\rm even}\ra$ is an odd parity state.

Then the protocol runs as in the $(2,2)$ case with the obvious
generalization of the carrier and the uploading and downloading
operators. In fact in both types of rounds the carrier can be
written as
\begin{equation}\label{carrierN}
  |\phi\ra=\frac{1}{\sqrt{2}}\sum_{q\in Z_2}|q,\overline{q}\ra,
\end{equation}
where $|\overline{q}\ra$ stands for the encoding in
(\ref{encodingnn}) and we have suppressed the subscripts "even"
and "odd" for simplicity. The uploading operator will be ${\cal
C}_A:=C_{a,1}C_{a,2}\cdots C_{a,n}$ for the odd rounds and ${\cal
C}_{A}:= C_{a,1}$ for the even rounds. The downloading operator
will be the same for both rounds and will be ${\cal
C}_B:=C_{b_1,1}C_{b_2,2}\cdots C_{b_n,n}$. \\

To show that the protocol runs in exactly the same way as in the
(2,2) scheme, we need to prove the basic properties of the
encoded states and the carrier. To this end, we first note from
(\ref{encodingnn}) that the following relations hold,
\begin{equation}\label{evenoddH}
H^{\otimes n}|\overline{s}_{\rm
odd}\ra=\frac{1}{\sqrt{2}}\sum_{x=0}^1 (-1)^{sx}
|\overline{x}_{\rm even }\ra,\h H^{\otimes n}|\overline{s}_{\rm
even}\ra=\frac{1}{\sqrt{2}}\sum_{x=0}^1 (-1)^{sx}
|\overline{x}_{\rm odd }\ra.
\end{equation}

From these two relations one easily shows that the carriers in
the even and odd rounds are turned into each other by the local
Hadamard actions of players. Second we need the generalization of
the properties (\ref{identitiesodd} and \ref{identitieseven}) to
the (n,n) case. To this end we start from the simple properties
\begin{equation}\label{uploadnn}
  C_{a,1}|q\ra_a|+\ra_{1}=|q\ra_a|+\ra_{1}\h
  C_{a,1}|q\ra_a|-\ra_{1}=(-1)^q|q\ra_a|-\ra_{1},
  \end{equation}
to obtain
\begin{eqnarray}\label{uploadevenN}
  C_{a,1}|q\ra_a |\overline{s}_{\rm even}\ra_{1,\cdots n}&=&\frac{1}{\sqrt{2}}
  \left( C_{a,1}|q\ra_a (|+,+,\cdots +\ra + (-1)^s |-,-,\cdots -\ra)_{1,\cdots n}\right)\cr
  &=& \frac{1}{\sqrt{2}}\left(|q\ra_a (|+,+,\cdots +\ra + (-1)^{s+q} |-,-,\cdots -\ra)_{1,\cdots
  n}\right)\cr &=&|q\ra_{a}|\overline{(q+s)}_{\rm even}\ra_{1,\cdots
  n}.
  \end{eqnarray}
The only other non-trivial relation which we should prove is the
following relation for the even rounds which is necessary for the
downloading operation, (for the odd rounds, the involved states
are product states and the relation is obvious):
\begin{equation}\label{nontrival}
  C_{b_1,1}C_{b_2,2}\cdots
  C_{b_n,n}|\overline{q}\ra|\overline{s}\ra
\end{equation}

To show the validity of this relation we first use the following
simple property of CNOT operation, where the first bit is the
control and the second bit is the target qubits:
\begin{equation}\label{cnotproperties}
  C|+,+\ra=|+,+\ra, \ \  C|+,-\ra=|-,-\ra,\ \ C|-,+\ra=|-,+\ra,\ \ C|-,-\ra=|+,-\ra.
\end{equation}
Second we use these properties and (\ref{encodingnn}) and the
abbreviation ${\cal C}_B:=C_{b_1,1}C_{b_2,2}\cdots
  C_{b_n,n}$ to obtain
\begin{eqnarray}\label{nontrival}
  {\cal C}_B|\overline{q}\ra|\overline{s}\ra &=& \frac{1}{2}{\cal
  C}_B\big(|+\ra^{\otimes n}+(-1)^q |-\ra^{\otimes n}\big)\big(|+\ra^{\otimes n}+(-1)^s |-\ra^{\otimes
  n}\big)\cr &=&
\frac{1}{2}\big(|+\ra^{\otimes n}|+\ra^{\otimes n}+
(-1)^q|-\ra^{\otimes n}|+\ra^{\otimes n} \cr &+&
(-1)^s|-\ra^{\otimes n}|-\ra^{\otimes n}+(-1)^{q+s}|+\ra^{\otimes
n}|-\ra^{\otimes n}\big)\cr &=& \frac{1}{2}\big(|+\ra^{\otimes
n}+(-1)^q |-\ra^{\otimes n}\big)\big(|+\ra^{\otimes n}+(-1)^{q+s}
|-\ra^{\otimes
  n}\big)=|\overline{q}\ra|\overline{q+s}\ra.
\end{eqnarray}
This completes the description and validity of the uploading and
downloading procedures for the (n,n) scheme.\\

In passing we note that the form of the carrier (\ref{carrierN})
for this $(n,n)$ secret sharing scheme is the same as in the
simplest cryptographic protocol,(\ref{(1,1)carrier}). We will see
in the next section that the appropriate carrier for the threshold
scheme $(k,n)$ where $n$ is an odd prime, is of the same form. We
will explain the reason for this general structure in the last
section, however before that, we explain in detail the carrier
for the $(k,n)$ secret sharing scheme.

\section{Quantum Carrier For $(k,n)$ Threshold Secret
Sharing}\label{kncarrier} There are situations where there are $n$
players and any subset of $k$ or more members can retrieve the
secret, while subsets of smaller size cannot. This is called a
$(k,n)$ threshold structure \cite{blakely,sham} in which all the
players have equal weight. One can also imagine situations where
different players have different weights. This leads to a general
access structure, according to which the players form a set
${\cal R}$ of say $n$ members and an access structure is a
collection ${\cal A}  $ of subsets of ${\cal R}$. The subsets in
${\cal A}$ (and their unions) are called authorized subsets and
the members of each authorized subset should be able to retrieve
the key by their collaboration, while the subsets which are not
in ${\cal A}$, called adversaries, cannot retrieve the secret. It
is known that once a threshold scheme is solved, then other more
general access structures will be possible \cite{Milad1,Milad2}.
For example if ${\cal R}=\{a,b,c\}$ and ${\cal
A}=\{\{a,b\},\{a,c\}\}$ , then we can run a $(3,4)$ threshold
scheme giving 2 shares to $a$
and one shares to $b$ and $c$ each. \\

The $(k,n)$ threshold scheme was first generalized to the quantum
domain in \cite{cleve}, where quantum states could also be shared
between $n$ parties so that any $k$ of the players could retrieve
the quantum state collaboratively. To be in conformity with the
no-cloning theorem, $n$ had to be smaller than $2k$. We will deal
in detail with the case where $n=2k-1$ is a prime number. Other
cases where $n<2k-1$ are obtained by a simple modification of the
$(k,2k-1)$ scheme. For example a scheme like $(k,2k-m)$ is
implemented by running the scheme $(k,2k-1)$ as usual, but with
Alice playing the role of the other $m-1$ receivers in addition
to her usual role. The idea of \cite{cleve} was to adapt the
polynomial code, first developed in \cite{aha}, to the quantum
domain. Note that in \cite{cleve}, quantum mechanics was
exploited only for message splitting and not for message
distribution. Later it was shown in \cite{bar1,bar2} that graph
states can be used for combining the two parts of the problem in
one scheme, for some threshold schemes, namely for $(2,3)$,
$(3,5)$ and $(n,n)$ schemes. Here we show that the idea of quantum
carrier can be used to provide a method of secure distribution
for all secrets of the $(k,n)$ types provided in \cite{cleve}. Let
us first see what a polynomial code is.

\subsection{The polynomial code}\label{polycode}

Consider a symbol $s$. Classically if we want to share this
symbol as a secret between $n$ parties, called $B_1$, $B_2,
\cdots B_n$, so that any $k$ members of the parties can retrieve
this symbol and fewer than $k$ members cannot, we can define a
real polynomial of degree $k-1$ in the form
\begin{equation}\label{poly}
  P_{{\bf c},s}(x):=c_0+c_1\ x + c_2\ x^2 + \cdots c_{k-2} x^{k-2}+ s\ x^{k-1}
\end{equation}
and evaluate this polynomial on $n$ distinct points $x_0,\cdots
x_{n-2} $, and  $x_{n-1}$. We can then give the member $B_i$ of
the set, the value $P_{{\bf c},s}(x_i)$. It is a simple fact that
a polynomial of degree $k-1$ is completely determined by its
values on $k$ distinct points. So any $k$ members can compare
their values and determine the full functional form of the
polynomial and hence the real number $s$. To make the process
simple and less prone to errors, we can substitute the real
number field with the field $Z_n:=\{0,1,2,\cdots n-1\}$ (where
$n$ is prime). For the $n$ points in $Z_n$ we can take simply
$x_i=i$. Hence we can encode the symbol $s$ into a product state
$|s\ra:=|P_{{\bf c},s }(0)\ra_{_{B_1}}\otimes |P_{{\bf c},s
}(1)\ra_{_{B_2}}\cdots \otimes |P_{{\bf c},s }(n-1)\ra_{_{B_n}}$.
Let us now sum such a product state over all possible ${\bf c}\in
Z_n^{k-1}$, and obtain the code
\begin{eqnarray}\label{encodeS}
  s\lo |\overline{s}\ra&:=& \frac{1}{\sqrt{n^{k-1}}}\sum_{{\bf c}\in Z_n^{k-1}}|P_{{\bf
c},s }(0)\ra_{_{B_1}}\otimes |P_{{\bf c},s }(1)\ra_{_{B_2}}\cdots
\otimes |P_{{\bf c},s }(n-1)\ra_{_{B_n}}  \cr &\equiv &
\frac{1}{\sqrt{n^{k-1}}}\sum_{{\bf c}\in Z_n^{k-1}}|P_{{\bf c},s
}(0), P_{{\bf c},s }(1), \cdots , \cdots P_{{\bf c},s }(n-1)\ra.
\end{eqnarray}
In order to see how to find a suitable carrier for this code, and
indeed in order to show that the carrier for this code falls
within the same class of carriers considered so far, we have to
prove further algebraic properties of this code. To do this, we
cast it in the form of a Calderbank-Shor-Steane (CSS) code
\cite{cs,ste}.

\subsection{The CSS structure of the polynomial
code}\label{csscode}

Let $n$ be a prime number. With addition and multiplication
modulo $n$, the set $Z_n:=\{0,1,2\cdots n-1\}$ will be a field.
For any $n$, $Z_n^{n}$ is a a vector space over $Z_n$, i.e.
$Z_n^{n}$ is the set of all $n$-tuples $(v_1, v_2, \cdots v_{n})$
where $v_i\in Z_n$. Let $C$ be a linear code, i.e. a subspace of
$Z_n^n$, spanned by linearly independent vectors $\{{\bf
e}_0,{\bf e}_1,\cdots {\bf e}_{k-2}\}$. Thus $C$ is isomorphic to
$Z_n^{k-1}.$ Consider the case where the dual code of $C$, i.e.
the code space spanned by all the vectors which are perpendicular
to $C$, contains $C$ and has one more dimension. Let $C^{\perp}$
be spanned by the vectors $\{{\bf e}_0,{\bf e}_1,\cdots, {\bf
e}_{k-2}\}\bigcup \{{\bf e}_{k-1}\}$. Thus we have
\begin{eqnarray}\label{ccperp}
C &=& {\rm Span} \{{\bf e}_0, {\bf e}_1, \cdots {\bf
e}_{k-2}\}\nn \\
C^{\perp} &=& {\rm Span} \{{\bf e}_0, {\bf e}_1, \cdots {\bf
e}_{k-2}, {\bf e}_{k-1}\}.
\end{eqnarray}
In the codes that we will introduce, the special vector ${\bf
e}:={\bf e}_{k-1}$ is normalized so that ${\bf e}\cdot {\bf
e}=-1.$ We therefore have
\begin{eqnarray}\label{eiej}
&&{\bf e}_i\cdot {\bf e}_j=0, \h \ \ \ \ 1\leq i,j\leq k-2, \nonumber\\
&&{\bf e}\cdot {\bf e}_j=0, \h \ \ \ \ \ 0\leq j\leq k-2, \\
&&{\bf e}\cdot {\bf e}=-1.\nn \ \ \ \
\end{eqnarray} We will now
define the following special Chalderbank-Steane-Shor (CSS) code
\cite{cs,ste,bch,sarvepalli}, whose codewords correspond to the
classes of the quotient space $C^{\perp}/C$:
\begin{equation}\label{codewords}
  |\overline{s}\ra:=\frac{1}{\sqrt{n^{k-1}}}\sum_{{\bf c}\in
  C}|{\bf c}+s {\bf e}\ra\
\end{equation}
Thus one dit $s$ is coded into the $n$-qudit state
$|\overline{s}\ra$ which is to be distributed between the $n$
receivers, each component of the vector being given to one participant.\\

In the appendix, we will show the vectors $e_l$ with components
\begin{equation}\label{ecompact}
  \left({\bf e}_l\right)_j=j^{l}, \ \ \ j=0,\ 1,\cdots n-1,\h
  l=0,\ 1,\cdots k-1
\end{equation}
satisfy all the properties listed in (\ref{eiej}). Explicitly we
have
\begin{eqnarray}\label{ecomponents}
  {\bf e}_0&=&(1,\ 1,\ 1,\  \cdots\cdots 1),\cr
   {\bf e}_{l\geq 1}&=&(0,\ 1,\ 2^l,\cdots (n-1)^l).
  \end{eqnarray}
It is now very simple to see that the CSS code thus constructed
is nothing but the polynomial code in (\ref{encodeS}). To see
this we note that the vector ${\bf c}\in C$ has the following
expansion
\begin{equation}\label{exp}
  {\bf c}=\sum_{l=0}^{k-2}{c_l}e_l
\end{equation}
and hence the components will be
\begin{equation}\label{exp}
  ({\bf c}+s{\bf e})_j=\left(\sum_{l=0}^{k-2}{c_l}({\bf e}_l)_j\right)+s({\bf
  e}_{k-1})_j=(\sum_{l=0}^{k-2}{c_l}j^l)+sj^{k-1}=P_{{\bf c},s}(j).
\end{equation}
Therefore giving each components to one player, the code will be
\begin{equation}\label{codewords}
  |\overline{s}\ra_{b_1, b_2, \cdots b_n}:=\frac{1}{\sqrt{n^{k-1}}}\sum_{{\bf c}\in
  Z_n^{k-1}}|P_{{\bf c},s}(0), P_{{\bf c},s}(1), \cdots, P_{{\bf
  c},s}(n-1)\ra,
\end{equation}
which is exactly the polynomial code (\ref{poly}). Now that the
CSS structure of the polynomial code is revealed, many of its
properties can be proved in a simple way. In particular we need the following property which plays an important role in the security of the carrier.\\

\textbf{Lemma:}{\it The set of all $(k,n)$ codes (\ref{codewords}),
is invariant under the joint multi-local Hadamard operation, i.e.
\begin{equation}\label{Hadamard}
  H^{\otimes n}|\overline{s}\ra =
  \frac{1}{\sqrt{n}}\sum_{x=0}^{n-1}\omega^{-sx}|\overline{x}\ra,
\end{equation}
where $\omega$ is a root of unity, $\omega^n=1$.\\}

We use a well-known property of the CSS codes according to which,
\begin{equation}\label{arrive2}
  H^{\otimes n}|\overline{\bf w}\ra= \frac{1}{|C_1/C_2|}\sum_{\overline{\bf
  v}}\omega^{{\bf w}\cdot {\bf v}}|\overline{\bf v}\ra,
\end{equation}
where $|C_1/C_2|$ is the number of cosets $C_1/C_2$. To adapt
this general relation to our case, we note that in our case
$|C^1|=|C^\perp|=n^k$, $|C^2|=|C|=n^{k-1}$, and hence
$|C^1/C^2|=n$. Moreover we make the following substitutions,
\begin{equation}\label{subs}
  |{\bf w}\ra\lo |{\bf c}+s{\bf e}\ra=|\overline{s}\ra,\ \ \ |{\bf v}\ra\lo |{\bf c}+x{\bf e}\ra=|\overline{x}\ra,\ \ \
\end{equation}
and note that
\begin{equation}\label{dot}
  ({\bf w})\cdot({\bf v})=({\bf c}+s{\bf e})\cdot({\bf c}+x{\bf
  e})=-sx.
\end{equation}
Putting all this together proves the lemma. \\
\subsection{The carrier and the uploading and downloading
operators}\label{uploaddownload}

The quantum carrier is constructed as follows:
\begin{equation}\label{carrier2}
  |\phi\ra_{_{a,b_1,\cdots,b_n}}=\frac{1}{\sqrt{n}}\sum_{q\in Z_n}|q\ra_{_a}|\overline{q}\ra_{_{b_1,\cdots
  b_n}}.
\end{equation}
The important property of this carrier is that it is invariant
under the joint action of Hadamard operators, performed by Alice
and all the other players. Using (\ref{Hadamard}) proves this
assertion:
\begin{equation}\label{Hn+1}
 H^{\otimes n+1}|\phi\ra =
 \frac{1}{n}\sum_{x,y,s}\omega^{sx-sy}|x,\overline{y}\ra=\frac{1}{\sqrt{n}}\sum_{x}|x,\overline{x}\ra
 =|\phi\ra.
 \end{equation}
In order to see how Alice uploads secrets onto the carrier and
how the players download the secret from the carrier we need some
algebraic properties of the code.\\

\textbf{Definition:} For any vector ${\bf v}=(v_1, v_2, \cdots
v_n)\in Z_n^n$ define the following string of CNOT operators
performed by Alice:
\begin{equation}\label{uppoly}
  {\cal C}_{A}({\bf v}):= C^{v_1}_{a,1}C^{v_2}_{a,2}\cdots
  C^{v_n}_{a,n}.
\end{equation}
Also define the following multi-local operator for Bob's:
\begin{equation}\label{downpoly}
  {\cal C}_{B}:= C_{b_1,1}C_{b_2,2}\cdots
  C_{b_n,n}.
\end{equation}

\textbf{Theorem:} {\it  The operator ${\cal C}_{A}:={\cal
C}_A({\bf e})$, for ${\bf e}$ as in (\ref{ecomponents}), uploads
the message into the carrier by Alice and the operator ${\cal
C}^{-1}_B$ downloads the message from the carrier by Bob's,
leaving the carrier in its original form.}\\

We fist show that for any state $|q\ra_a$ and any
message $|\overline{s}\ra_{1,2,\cdots n}$
\begin{equation}\label{proofA}
{\cal C}_{A}|q\ra|\overline{s}\ra=|q\ra|\overline{s+q}\ra.
\end{equation}
This is seen by expansion of $|\overline{s}\ra$ in components and
noting that
\begin{eqnarray}\label{proofA2}
{\cal C}_{A}|q\ra|\overline{s}\ra &=&
\frac{1}{\sqrt{n^{k-1}}}{\cal C}_{A}|q\ra\sum_{{\bf c\in
Z_n^{k-1}} }|({\bf c}+s{\bf e})_1, ({\bf c}+s{\bf e})_2, \cdots
({\bf c}+s{\bf e})_n\ra\cr &=&
\frac{1}{\sqrt{n^{k-1}}}|q\ra\sum_{{\bf c \in Z_n^{k-1}} }|({\bf
c}+(s+q){\bf e})_1, ({\bf c}+(s+q){\bf e} )_2, \cdots ({\bf
c}+(s+q){\bf e })_n\ra\cr &=& |q\ra|\overline{s+q}\ra.
\end{eqnarray}
From (\ref{proofA2}), we see that
\begin{equation}\label{uploadcarrier}
  {\cal C}_A |\phi\ra|\overline{s}\ra = \frac{1}{\sqrt{n}}\sum_{q\in Z_n} {\cal C}_A |q,\overline{q}\ra
  |\overline{s}\ra = \frac{1}{\sqrt{n}}\sum_q |q,\overline{q}\ra
  |\overline{q+s}\ra.
\end{equation}
Therefore Alice uploads (entangles) the message
$|\overline{s}\ra$ to the carrier by the local operation ${\cal
C}_A$. For the other part, we need the following
\begin{equation}\label{proofB1}
{\cal
C}_{B}|\overline{q}\ra|\overline{s}\ra=|\overline{q}\ra|\overline{s+q}\ra.
\end{equation}
To show this we note that
\begin{eqnarray}\label{proofB2}
{\cal C}_{B}|\overline{q}\ra|\overline{s}\ra &=&{\cal
C}_B\sum_{{\bf c},{\bf c'}} |{\bf c}+q {\bf e}\ra |{\bf c'}+s{\bf
e}\ra \cr &=& \sum_{{\bf c},{\bf c'}} |{\bf c}+q {\bf e}\ra |{\bf
c'}+{\bf c} +(s+q){\bf e}\ra= |\overline{q}\ra|\overline{s+q}\ra.
\end{eqnarray}

From this last equation we find that
\begin{equation}\label{proofB3}
 {\cal C}^{-1}_B \sum_{q} |q,\overline{q}\ra
  |\overline{s+q}\ra = \sum_q |q,\overline{q}\ra
  |\overline{s}\ra,
\end{equation}
which means that the players, can download the message from the
carrier and put the carrier back to its original form. \\

The basic steps of the quantum secret sharing are now clear. A
carrier in the form of the state $|\phi\ra$ is shared between
Alice and all the receivers, $B_1, B_2, \cdots B_n$. Alice
operates by his ${\cal C}_{A}$ operator on her part of the
carrier and the code state $|\overline{s}\ra$ and thus entangles
the code state to the carrier $|\phi\ra$. At the destination the
players act on the carrier and the code space by ${\cal
C}_B^{-1}$ and download the state $|\overline{s}\ra$. From this
code state, no less than $k$ players can retrieve the secret
symbol $s$. The carrier is now ready for transferring the next
code state.

\section{The Security of the Quantum Carrier}\label{security}

In this section we discuss the security of state transmission via
the carrier and analyze two types of attacks performed by Eve.
The security of the retrieval procedure of the symbol $s$ from
the encoded state $|\overline{s}\ra$ need not concern us and has
been discussed elsewhere \cite{cleve}. Obviously the analysis of
security depends on the resources available to Eve. We consider
two types of attacks in the following two subsections. This type
of analysis applies to all the schemes mentioned up to now.

\subsection{Simple intercept of message by Eve, without her entanglement to the carrier}
In this type of attack we assume that Eve is not entangled with
the carrier, but she has access to all the message qudits sent
from Alice to the players. After uploading the message, the full
state is given by
\begin{equation}\label{psi''}
  |\phi_s\ra=\frac{1}{\sqrt{n}}\sum_{q\in Z_n}|q,\overline{q},\overline{q+s}\ra_{_{A,{{\bf
  B}},1\cdots n}}.
\end{equation}
While in transit the data qudits are in the state
\begin{equation}\label{data}
  \rho_D = \frac{1}{n}\sum_{s\in Z_n}|\overline{s}\ra\la \overline{s}|,
\end{equation}
which is an equal mixture of all the encoded states. Therefore
even if Eve have access to all the data in transit and intercepts
all the qudits sent to all the players, she cannot acquire the
secret $s$ or the secret state, since she
only finds and equal mixture of all the encoded states.\\

At the destination, the receivers, act by the inverse local
operator ${\cal C}_{B}$ and according to (\ref{proofB3}),
disentangle the code from the carrier. They can then retrieve the
classical secret $s$ by collaboration of each other according to
the access structure.  Once retrieved, we resort to the arguments
of \cite{cleve} to show that this encoded state is secure against
cheating of groups of
unauthorized players.  \\

Therefore in the simplest intercept attack, Eve does not acquire
any information about the secret symbol $s$. We now consider more
general attacks.

\subsection{Entanglement of Eve to the carrier and intercept-resend attack}
We now assume that in addition to access to the message channel,
Eve can entangle herself to the carrier. Let us see if she can do
appropriate action for intercepting the encoded state
$|\overline{s}\ra$ and not an equal mixture. Consider the first
round where the symbol $s_1$ is encoded to $|\overline{s}_1\ra$
and sent by Alice. The state of the carrier and the message after
Alice uploading operation will be

\begin{equation}\label{psi1}
  |\phi^{(1)}(s_1)\ra_{a,B,M}=\frac{1}{\sqrt{n}}\sum_{q}|q,\overline{q}\ra_{a,B}|\overline{q+s_1}\ra_M,
\end{equation}
where $a$ stands for Alice, $B$ for all the players $B_1, \cdots
B_n$ and $M$ for the $n$ message qudits, $m_1, m_2, \cdots m_n$.
Eve can now set her $n$ ancilla qudits $E:=(e_1, e_2, \cdots
e_n)$ to $|\overline{0}\ra_E$, and then do the following
operations:  acts by ${\cal
C}_{M,E}:=C_{m_1,e_1}C_{m_2,e_2}\cdots C_{m_n,e_n}$ which
transforms the state
$|\phi^{(1)}(s_1)\ra_{a,B,M}|\overline{0}\ra_E$ to
\begin{equation}\label{psi1}
  |\phi^{(2)}(s_1)\ra_{a,B,M,E}=\frac{1}{\sqrt{n}}\sum_{q}|q,\overline{q}\ra_{a,B}|\overline{q+s_1}\ra_M|\overline{q+s_1}\ra_E.
\end{equation}
When Alice and the players execute the first round of the
protocol to the end and the players extract the state
$|\overline{s_1}\ra$, Eve acquires nothing from the symbol $s_1$,
however she has achieved in entangling herself with the carrier
in the form

\begin{equation}\label{psi1}
  |\phi'\ra_{a,B,E}=\frac{1}{\sqrt{n}}\sum_{q}|q,\overline{q}\ra_{a,B}|\overline{q+s_1}\ra_E.
\end{equation}
In the second round, when the symbol $s_2$ is being sent and the
full state is of the form

\begin{equation}\label{psi2}
  |\phi'(s_2)\ra_{a,B,E,M}=\frac{1}{\sqrt{n}}\sum_{q}|q,\overline{q}\ra_{a,B}|\overline{q+s_2}\ra_M  |\overline{q+s_1}\ra_E,
\end{equation}
Eve performs the following sequence of operations: 1) acts by
$C^{-1}_{M,E}$ to produce the state
\begin{equation}\label{psi3}
  |\phi'^{(1)}(s_2)\ra_{a,B,E,M}=\frac{1}{\sqrt{n}}\sum_{q}|q,\overline{q}\ra_{a,B}|\overline{q+s_2}\ra_M
  |\overline{s_1-s_2}\ra_E,
\end{equation}
2) measures the ancillas to acquire $s_1-s_2$, and 3) acts by
${\cal C}_{M,E}$ to put back the full state in the form
(\ref{psi2}). When Alice and the players finish the second round,
they acquire the symbol $s_2$, but Eve also acquires the symbol
$s_1-s_2$, while she is still entangled with the carrier in the
form (\ref{psi1}) and is ready to do the same attack for the next
round. In this way she is able to retrieve the sequence of symbols
\begin{equation}\label{seq}
  s_1-s_2,\   s_1-s_3,\  s_1-s_4,\cdots.
\end{equation}
This sequence enables her to find the whole message by checking
$n$ different choices for the original symbol $s_1$. This shows
that if there is a possibility for Eve's entanglement with the
carrier, she is able to successfully intercept all the data. \\

In order to prevent this, Alice and the players act on their
respective qudits of the carrier, by Hadamard operations. As we
have seen above, this operation leaves the pure form of the
carrier invariant. Let us see if this operation is able to detect
an entanglement of Eve, i.e. a contamination of the carrier. It
is clear that if the carrier contains terms which are not of the
form $|q,\overline{q}\ra_{a,B}$, then there will be mismatches
between what Alice uploads and what the players download. This
mismatch can easily be detected by public announcements of some
stray bits which are deliberately inserted into the stream of the
symbols. In order to escape this detection, the only admissible
form of Eve's entanglement with the carrier is
\begin{equation}\label{sec-2}
  |\phi'\ra=\sum_{q}|q,\overline{q}\ra\otimes |\xi_q\ra,
\end{equation}
where $\xi_q$ are a collection of un-normalized states of Eve.
The method of detection in this case is the same as in the simple
$(2,2)$ case, discussed after equation (\ref{abe}). Also in view
of the discussion in subsection \ref{rollle}, only the legitimate
parties or even a subset of them are required to publicly
announce the results of their measurements.\\

In order to prevent this type of apparently undetectable
entanglement, we note that the pure carrier is invariant under
the action of Hadamard operators, while this contaminated carrier
is not. In order to retain the correlations, Eve may operate on
her ancilla by a suitable operator $U$ to change the above state
into
\begin{equation}\label{sec-1}
  (H\otimes H^{\otimes n}\otimes U)|\phi'\ra=\frac{1}{n}\sum_{q,x,y}\omega^{q(x-y)}|x,\overline{y}\ra\otimes
  U|\xi_q\ra.
\end{equation}
In order to retain the original form of correlations between
Alice and the players in the carrier, the operator $U$ must
satisfy the following property
\begin{equation}\label{sec-0}
  U\sum_{q}\omega^{q(x-y)}|\xi_q\ra=n|\eta_x\ra\delta_{x,y}\h \forall
  \ q,
\end{equation}
for some states $|\eta_x\ra$. Putting $x=y$ one finds that
$|\eta_x\ra$ is independent of $x$ and hence a rearrangement
shows that
\begin{equation}\label{sec}
  \sum_{q}\omega^{q(x-y)} (U|\xi_q\ra-|\eta\ra)=0.
\end{equation}
Acting on the left hand side by the inverse Hadamard operation,
one finds that $U(|\xi_q\ra -|\eta\ra)=0$, which means that all
the states $\xi_q$ are equal to each other and hence the state
$|\phi'\ra$ cannot be an entangled state. Therefore Eve cannot
entangle
herself to the Carrier without being detected. \\

Note that we have assumed that Eve is an external agent and all
the players have run the protocol as they should, i.e. have
performed their Hadamard operation on the carrier at the end of
each round. In principle one can assume that a subgroup of
players collaborate with Eve, i.e. perform other operations than
Hadamard in order to ease undetectable Entanglement of Eve with
the carrier. For example consider a $(2,3)$ scheme with players
$B_1$, $B_2$ and $B_3$. The question is whether one of the
players, say $B_1$ is capable to collaborate with Eve to retrieve
the secret symbol $s$? To be honest, we have not been able
neither to devise a successful attack of this type nor a method
for its prevention.

\subsection{The role of players collaborations}\label{rollle}
It is part of the protocol that all the players should perform
their CNOT's on the state $|\phi_s\ra$ in order to download the
state $|\overline{s}\ra$, but once this state is downloaded, then
no less than $k$ players can collaborate to retrieve the symbol
$s$ as proved in \cite{cleve}. One may then argue that this is
not a genuine $(k,n)$ threshold scheme, since for downloading the
state $|\overline{s}\ra$ from the carrier, all the players should
collaborate. \\

The important point is that the CNOT operation of all the players
are needed only for cleaning of the carrier from remnants of
messages, and in fact any $k$ players can retrieve the symbol $s$
from the state that they download from the carrier, but the
running of the protocol for other rounds, needs collaboration of
all the
players. \\

This assertion can be proved as follows: Let $K$ be any set of
$k$ members who want to retrieve the message. Denote by ${\bf
{\cal C}}_{K}$ the joint CNOT operations of the $k$ players
belonging to the set $K$, i.e. \be {\bf {\cal C}}_{{\bf
K}}:=\bigotimes_{j\in K}{\cal C}_{B_j,j}\ee Denote also by ${\bf
{\cal C}}_{{\bf N-K}}$ the joint CNOT operations of the rest of
$N-k$ players. Also denote by ${\cal M}_{\bf K}$, the local
operation and classical communications that the $k$ players
perform among themselves to recover the symbol $s$ from
$|\overline{s}\ra$.  We have shown that the sequence of
operations ${\cal M}_{\bf K}{\bf {\cal C}}_{\bf K}{\bf {\cal
C}}_{\bf N-K}$ when acting on the state $|\phi_s\ra$ in
(\ref{psi''}) produces the symbol $s$ unambiguously and leaves
the carrier clean of the remnants of the message, i.e.
disentangle the state $|\overline{s}\ra$ from the carrier. It is
important to note that due to their local nature the two
operations ${\cal M}_{\bf K}$ and ${\bf {\cal C}}_{\bf N-K}$
commute, so that we have the identity \be {\cal M}_{\bf K}{\bf
{\cal C}}_{\bf K}{\bf {\cal C}}_{\bf N-K}={\bf {\cal C}}_{\bf
N-K}{\cal M}_{\bf K}{\bf {\cal C}}_{\bf K} = {\bf {\cal C}}_{\bf
N-K}\left({\cal M}_{\bf K}{\bf {\cal C}}_{\bf K}\right). \ee
However if the operation on the left hand side of this relation,
leaves the $k$ players in the set $K$ with an unambiguous symbol
$s$, we can conclude that the operations $\left({\cal M}_{\bf
K}{\bf {\cal C}}_{\bf K}\right)$ does the same thing, because the
remaining operation ${\bf {\cal C}}_{\bf N-K}$, by its local
nature, has no effect on the qudits retrieved by the set $K$. The
sole effect of ${\cal C}_{\bf N-K} $ is to disentangle completely
the carrier from the message state and make it ready for the next
round. Such collaboration is of course necessary for the
continuous running of the protocol like any other communication
task. Let us now study a simple example, in which we will also
see in explicit terms the above argument.

\section{Example: The $(2,3)$ Threshold Scheme}\label{23scheme}

\begin{figure}[t]
\centering
   \includegraphics[width=8cm,height=3cm,angle=0]{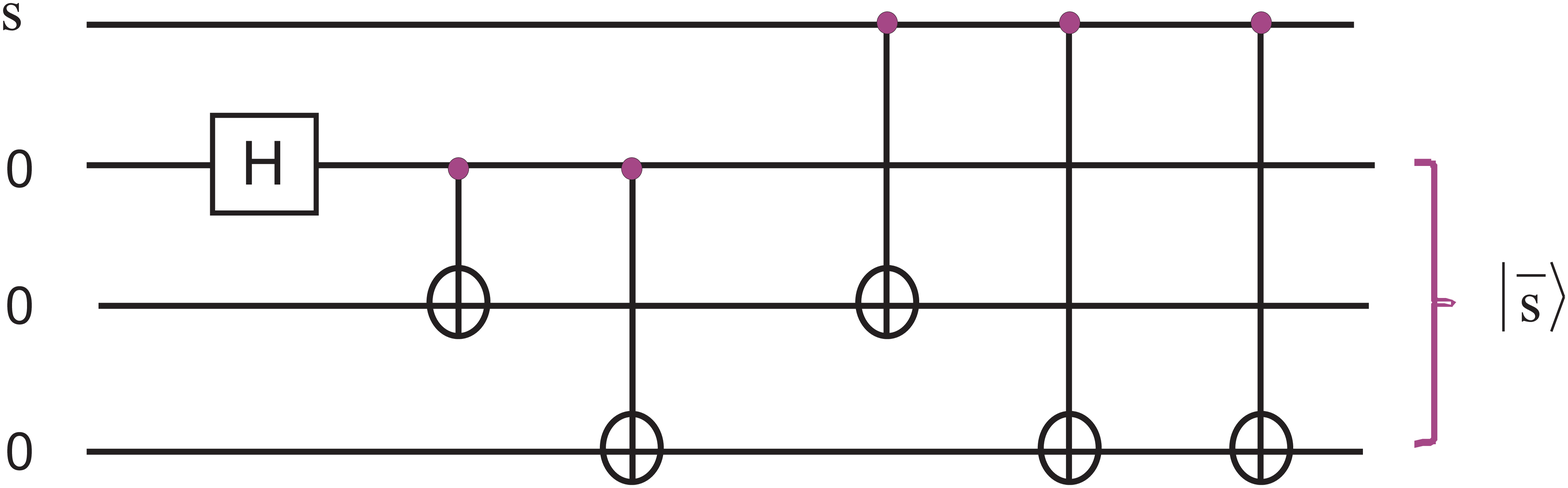}
   \caption{(Color Online) The encoding circuit used by Alice, for generating the (2,3) code $|\overline{s}\ra$ from
   $s$.
   } \label{circuit}
\end{figure}

\begin{figure}[t]
\centering
   \includegraphics[width=10cm,height=4cm,angle=0]{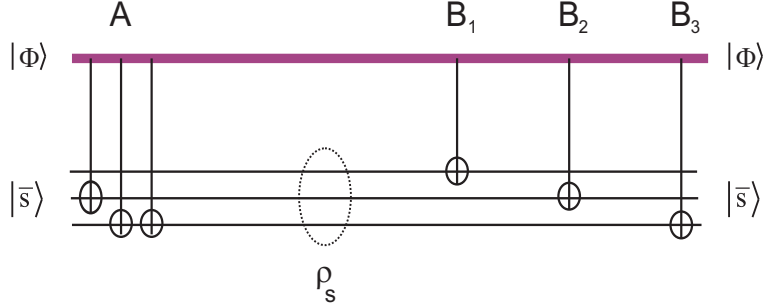}
   \caption{(Color Online) The schematic form of the carrier (thick line) and the uploading (by A) and downloading (by $B_1, B_2$ and $B_3$)
   for a simple $(2,3)$ threshold scheme.
   } \label{carrierfigure}
\end{figure}

The simplest threshold scheme is the scheme $(2,3)$ for which
\begin{equation}\label{pcs}
  P_{c,s}(x)=c_0+sx
\end{equation}
and hence
\begin{eqnarray}\label{35}
  s&\lo& \frac{1}{\sqrt{3}}\sum_{c_0}|c_0, c_0+s, c_0+2s,\ra\cr
  &=& (I\otimes X\otimes X^2)^s
  \frac{1}{\sqrt{3}}\sum_{c_0} |c_0, c_0, c_0\ra
\end{eqnarray}
or more explicitly as
\begin{eqnarray}\label{012}
  && 0 \rightarrow   |\bar{0} \rangle = \frac{1}{\sqrt{3}}\left(|000 \rangle + |111 \rangle + |222 \rangle\right) \nonumber \\
  && 1 \rightarrow   |\bar{1} \rangle =  \frac{1}{\sqrt{3}}\left(|012 \rangle + |120 \rangle + |201 \rangle\right)  \\
  && 2 \rightarrow   |\bar{2} \rangle =  \frac{1}{\sqrt{3}}\left(|021 \rangle + |102 \rangle + |210
    \rangle\right)\nn.
\end{eqnarray}
Note that for qudits, the operators $X$ and $Z$ (to be used
later) are defined as $X|i\ra=|i+1,\ \ \ {\rm mod} d\ra$ and
$Z|i\ra=\omega^i|i\ra$, where $\omega^d=1$. Equation (\ref{35})
shows what kind of encoding circuit Alice has to use to encode a
state $\sum_ s a_s |s\ra$ to $\sum_s a_s |\overline{s}\ra$. The
encoding circuit is shown in figure (\ref{circuit}).
Moreover it is easily seen from (\ref{012}) that the operator
$\overline{Z}$ defined as
\begin{equation}\label{zzz}
 \overline{Z}:= (I\otimes Z\otimes Z^2)
\end{equation}
acts as follows on the code states \be\label{propertyz}
\overline{Z}|\overline{s}\ra=\omega^{-s}|\overline{s}\ra. \ee
These encoded states have the nice properties that
\begin{eqnarray}\label{zz}
  Z_1^{-1}\otimes Z_2|\bar{s}\ra &=
   \omega^{s}|\bar{s}\ra,\nonumber\\
Z_2^{-1}\otimes Z_3|\bar{s}\ra &=
  \omega^{s}|\bar{s}\ra,\\
Z_3^{-1}\otimes Z_1|\bar{s}\ra &=
  \omega^{s}|\bar{s}\ra\nn,
\end{eqnarray}
which shows clearly that any two of the receivers can retrieve
the classical secret $s$ by local measurements of the encoded
state.  The uploading and downloading operators for this scheme
are shown in figure (\ref{carrierfigure}).\\

Let us now see explicitly in this example, an instance of the
general discussion after equation (\ref{sec}). In other words, we
want to show that although the CNOT action of all three receivers
is necessary for disentangling the state $|s\ra$ from the
carrier, it does not mean that full collaboration of the
participants is necessary for recovering the message $s$. That is
let us show that even without the collaboration of $B_3$, $B_1$
and $B_2$ can indeed disentangle and retrieve the message from
the carrier. The collaboration of $B_3$ is only needed to clean
the carrier from the message.\\

Assume that only two of the participants, say $B_1$ and $B_2$
enact their CNOT's on the state $|\phi_s\ra$. The resulting state
will be
\begin{equation}\label{psi3}\nn
{\cal C}^{-1}_{B_1}{\cal
C}^{-1}_{B_2}|\phi_s\rangle=\frac{1}{\sqrt{3}} \sum_{j,k}{|i
\rangle_{_{A}} |j,j+i,j+2i
\rangle_{_{B_1B_2B_3}}|k,k+s,k+j+2(i+s)\rangle_{123}}.
\end{equation}
Thus measurements of qudits $1$ and $2$ by these two participants
reveals the secret $s$, without any need for collaboration of
$B_3$. \\

On the other hand suppose that the player $B_3$ wants to retrieve
the message symbols on his own. To this end he adds to the
quantum carrier $|\phi\ra_{_{A,B_1,B_2,B_3}}$ an extra qudit, in
the state $|\xi_0\ra_{_{B'_3}}$ and at the end of each round,
when all the parties are supposed to act on the carrier by
Hadamard operators, $B_3$ acts by a suitable bi-local operator on
his two qudits, so that in conjunction with the Hadamard
operators of $A$, $B_1$ and $B_2$, the quantum carrier transforms
to

\begin{equation}\label{bob3A}
|\phi'\ra = \sum_{i,j}|i,j+i,j+2i,j\ra_{_{A,B_1,B_2,B_3}}
|\eta_j\ra_{_{B'_3}}
\end{equation}

This is the only operation he can do in order not to destroy the
correlations between stray qudits which is checked randomly by Alice and the participants. \\

Now when the other participants proceed as usual for entangling a
code state $|\overline{s}\ra$ to and from the quantum carrier,
$B_3$ wants to proceed in a different way to reveal the symbol
$s$ on his own. The state of the quantum carrier and the code
state $|\overline{s}\ra$, which at the beginning of a round is
$|\phi'\ra|\overline{s}\ra$, after Alice CNOT operations will
develop as follows:
\\
\begin{eqnarray}\label{step1}
|{\phi_{s}^{'}}\ra = \sum_{i,j}|i\ra_{_{A}}|j+i,j+2i,j,
\eta_j\ra_{_{B_1,B_2,B_3, B'_3}}|\overline{i+s}\ra_{1,2,3}
\end{eqnarray}
where we use the subscripts $1,2,3$ to denote the qudits which
are respectively sent to $B_1,B_2$ and $B_3$.\\

It is now easily verified that the density matrix of the qudits
$B_3, B'_3, 1, 2$ and $3$ is given by
\begin{equation}\label{rhobb}
  \rho_{_{B_3,B'_3,1,2,3}}=\sum_{j}|j\ra\la j|_{_{B_3}}\otimes
  |\eta_j\ra\la \eta_j|_{_{B'_3}}\otimes\sum_{i}|\overline{i+s}\ra\la
  \overline{i+s}|,
\end{equation}
which is independent of $s$. Therefore even when one of the
participants entangles a qudit to the quantum carrier, refrains
from cooperation with others in applying Hadamard gates and/or
inverse CNOT operations, he cannot obtain
any information about the secret symbol $s$.\\

The collaboration of all the participants, is only necessary for
disentangling completely the data from the carrier and making it
ready for next use. This is certainly a feature that any
communication protocol should have.

\section{Discussion}\label{diss}

We have developed the concept of quantum carrier \cite{zhang,bk}
to encompass more complex classical secret and quantum sharing
schemes. We have described the procedure of uploading and
downloading messages to and from the carrier in increasingly
complex situations, i.e. for quantum key distributions, for
$(2,2)$, $(n,n)$ and $(k,n)$ threshold schemes. As described in
the text, for each task a different quantum carrier is required,
although it seems that they all have similar forms
(\ref{carrier2}). We have also shown that simple intercept-resend
attacks can destroy the pattern of entanglement in the carrier
which can be detected by legitimate parties. In the general
$(k,n)$ secret sharing scheme, although collaboration of all
parties is required for the continuous running of the protocol
(i.e. cleaning of the carrier from the remnants of the
transmitted messages), any set of $k$ players can download and
retrieve the message.\\

We hope that together with the previous results, the concept of
quantum carrier can attract the attention of other researchers
who will develop it into more complex forms. An important
question is whether there can be universal carriers between a set
of players, which can be used for various cryptographic tasks on
demand of the players, i.e. quantum key distribution between the
sender and a particular receiver, or secret sharing between the
sender and a particular set of players.? Another interesting
general question is whether there exist general carriers which
can be used to simultaneously send many messages to different
receivers via a single quantum carrier, in the same way that
frequency modulation is used for such a goal in classical
communication. Finally the question of general proof of security
of these types of carrier-based protocols remain to be
investigated.

\section{Acknowledgement}

We would like to express our deep gratitude to the two referees
of this paper, specially referee B, whose careful reading of the
manuscript and many constructive suggestions were essential in
improvement of the presentation of our results. We also thank R.
Annabestani, S. Alipour, S. Baghbanzadeh, K. Gharavi, R.
Haghshenas, and A. Mani for very valuable comments. M. M. is
deeply indebted to M. R. Koochakie for very stimulating
discussions.  Finally V. K. would like to specially thank Farid
Karimipour, for his kind hospitality in Villa Paradiso, north of
Iran, where the major parts of this manuscript was written.
\section{Appendix}
In this appendix we want to prove that the vectors in
(\ref{ecomponents}) satisfy the properties (\ref{codewords}).
That is if we define
\begin{equation}\label{ecompactA}
  \left({\bf e}_l\right)_j=j^{l}, \ \ \ j=0,\ 1,\cdots n-1,\h
  l=0,\ 1,\cdots k-1,
\end{equation}
then
\begin{eqnarray}\label{eiejA}
&&{\bf e}_i\cdot {\bf e}_j=0, \h \ \ \ \ 1\leq i,j\leq k-2, \nonumber\\
&&{\bf e}\cdot {\bf e}_j=0, \h \ \ \ \ \ 0\leq j\leq k-2, \\
&&{\bf e}\cdot {\bf e}=-1.\nn \ \ \ \
\end{eqnarray}

Let $p$ be an odd prime and define
\begin{equation}\label{Sk}
  S_k(p):=\sum_{j=1}^{p-1}j^k.
\end{equation}
First we prove that,
\begin{equation}\label{Skresult}
  S_k(p)=-\delta_{k,p-1},\h {\rm mod} \ p.
\end{equation}

Consider the following identity:
\begin{equation}\label{identity}
  \sum_{j=1}^{p-1}\left[(j+1)^m-j^m\right]=p^m-1\equiv -1.
\end{equation}
Expand the first term by using the binomial theorem to find
\begin{equation}\label{identity2}
  \sum_{j=1}^{p-1}\left[1+\sum_{r=1}^{m-1}\left(\begin{array}{c} m \\
  r\end{array}\right)j^{r}\right]=-1.
\end{equation}
Interchange the order of the two summations and use the
definition (\ref{Sk}) to obtain

\begin{equation}\label{simp}
  \sum_{r=1}^{m-2}\left(\begin{array}{c} m \\
  r\end{array}\right)S_r(p)+mS_{m-1}(p)= 0.
\end{equation}
This gives us a recursion relation in the form

\begin{equation}\label{recur}
  S_{m-1}(p)=\frac{-1}{m}\sum_{r=1}^{m-2}\left(\begin{array}{c} m \\
  r\end{array}\right)S_r(p).
\end{equation}
The recursion relation is valid for $2 < m < p$. The lower bound
is obvious from the upper limit on the summation. The upper bound
is due to the fact that for $m=p$, the denominator itself
vanishes modulo $p$. Equation (\ref{recur}) leads for example to

\begin{eqnarray}\label{recurexample}
  S_2(p) &=& -\frac{1}{3}\left[\left(\begin{array}{c} 3 \\
  1\end{array}\right)S_1(p)\right]\nonumber \\
 S_3(p) &=& -\frac{1}{4}\left[\left(\begin{array}{c} 4 \\
  1\end{array}\right)S_1(p)+ \left(\begin{array}{c} 4 \\
  2\end{array}\right)S_2(p)\right] \\
 S_4(p) &=& -\frac{1}{5}\left[\left(\begin{array}{c} 5 \\
  1\end{array}\right)S_1(p)+ \left(\begin{array}{c} 5 \\
  2\end{array}\right)S_2(p)+ \left(\begin{array}{c} 5 \\
  3\end{array}\right)S_3(p)\right]\nonumber  \\
&\cdots& \nn
\end{eqnarray}
Direct calculation gives $S_1(p)=1+2+3+\cdots
(p-1)=\frac{p(p-1)}{2}$ which is zero mod $p$, since $p-1$ is
even. The recursion relations above then imply that $S_m(p)=0$
for all $1\leq m \ <\  p$. The case $m=p$ should be calculated
directly, using the Euler theorem which states that for every
prime number $p$, \ \ $j^{p-1}= 1 \ \ {\rm mod} \ \ p$. The
result is immediate, namely $S_{p-1}(p)=-1.$ \\

Now using the relation (\ref{Skresult}) it's easy to verify that
the vectors ${\bf e}_i$ in (\ref{ecompactA}) satisfy the desired
properties in (\ref{eiejA}).


\begin{thebibliography}{}

\bibitem{hbb} M. Hillery {\it et al}., {\it Phys. Rev. A} {\bf 59} (1999) 1829.
\bibitem{karlsson} A. Karlsson {\it et al}., {\it Phys. Rev. A} {\bf 59} (1999) 162.
\bibitem{tripartite} A. M. Lance {\it et al}., {\it Phys. Rev. Lett} {\bf 92} (2004) 177903.
\bibitem{tyc} T. Tyc and B. C. Sanders, {\it Phys. Rev. A} {\bf 65} (2002) 042310.
\bibitem{contisanders} A. M. Lance {\it et al}., {\it Phys. Rev. A} {\bf 71} (2005) 033814; A. M. Lance {\it et al}., {\it New. Journal of Physics} {\bf 5} (2003).

\bibitem{Exp1} Yu-Ao Chen {\it et al}., {\it Phys. Rev. Lett.} {\bf 95} (2005) 200502.
\bibitem{Exp2}  S. Gaertner {\it et al}., {\it Phys. Rev. Lett.} {\bf 98} (2007) 020503.
\bibitem{bk} S. Bagherinezhad and V. Karimipour,  {\it Phys. Rev. A} {\bf 67} (2003) 044302;
V. Karimipour, {\it Phys. Rev. A} {\bf 72} (2005) 056301; V.Karimipour, {\it Phys. Rev. A} {\bf 74} (2006) 016302.
\bibitem{zhang2} Zj. Zhang and Zx.  Man, {\it Phys. Rev. A} {\bf 72} (2005) 022303.
\bibitem{liy}  Zj. Zhang. Y. Li and Zx. Man, {\it Phys. Rev. A} {\bf 71} (2005) 044301.
\bibitem{repeaters1} H. Briegel {\it et al}., {\it Phys. Rev. Lett.} {\bf 81} (1991) 5932-5935.
\bibitem{repeaters2} Z. S. Yuan {\it et al}., {\it Nature} {\bf 454} (2008) 1098-1101.
\bibitem{repeaters3} K. F. Reim {\it et al}., {\it Phys. Rev. Lett.} {\bf 107} (2011) 053603.
\bibitem{zhang} Y. S. Zhang{\it et al}., {\it Phys. Rev. A} {\bf 64} (2001) 024302.
\bibitem{cleve} R. Cleve {\it et al}., {\it Phys. Rev. Lett.} {\bf 83} (1999) 648.
\bibitem{sarvepalli} P. K. Sarvepalli and A. Klappenecker, {\it Phys. Rev. A} {\bf 80} (2009) 022321.
\bibitem{bar1} D. Markham and B. C. Sanders, {\it Phys. Rev. A} {\bf 78} (2008) 042309.
\bibitem{bar2} A. Keet {\it et al}., {\it Phys. Rev. A} {\bf 82} (2010) 062315.
\bibitem{graph} M. Hein {\it et al}., {\it Proc. of the International School of Physics "Enrico Fermi"
on "Quantum Computers, Algorithms and Chaos"} Varenna, Italy, July, 2005.
\bibitem{blakely} G. Blakely, {\it Proc. AFIPS 48},1979, pp. 313-317.

\bibitem{sham} A. Shamir, {\it Commun. ACM 22}, 1979, pp. 612.

\bibitem{Milad1}S. Goldwasser, {\it  Proc. of CRYPTO 88}, Santa Barbara, 1990, vol. 403 of LNCS.

\bibitem{Milad2} D. Gottesman, {\it Phys. Rev. A} {\bf 61} (2000) 042311.

\bibitem{aha} D. Aharonov and M. Ben-Or, {\it Proc. 29th Ann. ACM Symp. on Theory of Computing},1998, pp. 176-188.
\bibitem{cs} A. R. Calderbank and P.W. Shor, {\it Phys. Rev. A} {\bf54}  (1996) 1098.
\bibitem{ste} A. Steane,  {\it Proc. Roy. Soc. Lond. A 452}, 1996, pp. 2551-2577.

\bibitem{bch} M. Grassl, T. Beth, {\it Proceedings X. International Symposium on Theoretical Electrical Engineering}, Magdeburg, 1999; M. Grassl {\it et al}., {\it International Journal of
Foundations of Computer Science (IJFCS)}, 2003, Vol. 14, No. 5, pp. 757-775.

\end{thebibliography}
\end{document}